\documentclass[11pt]{article}
\usepackage[dvips]{graphics,graphicx}
\DeclareGraphicsExtensions{.eps}

\usepackage{a4}
\def\ds{\displaystyle}

\def\id{\mbox{\small 1}\!\!1}

\def\nn{\nonumber}

\def\cal{\mathcal}

\def\bspace{\!\!\!\!\!\!\!\!\!\!}

\def\qed{\hbox{${\vcenter{\vbox{                        
   \hrule height 0.4pt\hbox{\vrule width 0.4pt height 6pt
   \kern5pt\vrule width 0.4pt}\hrule height 0.4pt}}}$}}

\def\R{\bf\sf R}
\def\Z{\bf\sf Z}
\def\C{\bf\sf C}

\newtheorem{lemma}{Lemma}

\newtheorem{definition}{Definition}

\begin{document}

\begin{center}
  
{\bf \Large Introduction to Pseudo-Group}

\bigskip

{\large S.C. Woon}

\medskip

Trinity College, University of Cambridge, Cambridge CB2 1TQ, UK

s.c.woon@damtp.cam.ac.uk


December 26, 1998

\bigskip

{\bf \large Abstract}

\end{center}

This paper introduces the idea of pseudo-group.  Applications of
pseudo-groups in Group Theory and Symmetry Breaking in Particle
Physics and Cosmology are considered.

\bigskip

\section{Definitions}

A group $\,G\,$ is a set with a binary operation which satisfies the
four group axioms: distributivity, associativity, existence of
inverse, and existence of identity.

\begin{definition}[{Pseudo-Group}]\quad\\
\label{d:pseudo-group}
A pseudo-group $\,G(\rho_1,\rho_2,\dots,\rho_k)\,$ of a group $\,G\,$
is a set with a binary operation which gradually acquires the group
properties of $\,G\,$ and gradually satisfies the group axioms as 
{\em some} of the parameters $\,\rho_1,\rho_2,\dots,\rho_k\,$ of the set
approach certain limiting values or tend asymptotically to infinity.
\end{definition}

We shall study how concrete representations of pseudo-groups can be
constructed, and consider their applications in Group Theory and
Symmetry Breaking in Particle Physics and Cosmology.

\section{Construction of Pseudo-group Representations}

A method of constructing pseudo-group representations is by means of
a variant of Fractional Derivative.

\subsection{Method of Fractional Derivative}

It is possible to classify Fractional Derivative into two types, each
supporting a different view or school of thought.

\begin{definition}[{{\em Type I \ Fractional Derivative}}]\quad\\
  Adopt the view that fractional derivative is an abstract
  analytic extension of ordinary derivative without invoking any
  physical picture. Fractional derivatives can be non-commutative,
  and the fractional derivative of a constant can be non-zero.
\end{definition}

For instance, the Riemann-Liouville Fractional Derivative
belongs to {\em Type I}\,.  The Riemann-Liouville Fractional
Derivative \cite{frac_calculus} begins with
\begin{equation}
\int_a^x f(\hat{x})\,(d\hat{x})^n\,\equiv\,
\underbrace{\int^{x}_{\!a}\!\int^{x_n}_{\!a}\!\!
  \int^{x_{n-1}}_{\!a}\!\!\!\!\!\!\!\!\cdots\!\int^{x_3}_{\!a}
  \!\!\int^{x_2}_{\!a}\!}_{n \mbox{\footnotesize-times}} f(x_1) \;
dx_1\,dx_2 \cdots dx_{n-1}\,dx_n
\end{equation}
for $n\in\Z+$ as the fundamental defining expression, and it can be
shown \cite[p. 38]{frac_calculus} to be equal to the Cauchy formula for
repeated integration,
\begin{equation}
\frac{1}{\Gamma{(n)}}
\int_a^{x}\!\!\frac{f(t)}{(x\!-\!t)^{1-n}}\,dt\;.\label{RL_int}
\end{equation}

The Riemann-Liouville fractional integral is analytically extended
from (\ref{RL_int}) as
\begin{eqnarray}
D^\sigma_{\!x|a} f(x)&=&\frac{d^\sigma}{dx^\sigma} f(x)
\,=\,\int_a^x f(x) (dx)^{-\sigma}\nn\\
&=&\frac{1}{\Gamma(-\sigma)}\int_a^x\!\!
\frac{f(t)}{(x\!-\!t)^{1+\sigma}}\;dt\quad(\sigma<0,\;\sigma,a\in\R)\;\;
\mbox{by } (\ref{RL_int})\;,\label{frac_int}
\end{eqnarray}
and the R-L fractional derivative is in turn derived from the
R-L fractional integral (\ref{frac_int}) by ordinary differentiation:
\begin{equation}
D^\sigma_{\!x|a} f(x)\,=\,D^m_{\!x|a} \left( D_{\!x|a}^{-(m-\sigma)}
  f(x) \right)\quad(\sigma>0,\;m\in\Z^+)
\label{e:Dxfm}
\end{equation}
where $m$ is chosen such that $m>1+\sigma,\; \sigma>0$.

\begin{lemma}
The equation (\ref{e:Dxfm}) is independent of the choice of $m$ for
$m>1\!+\!\sigma,\; m\in\Z^+,\; \sigma\in\R,\; \sigma>0$.
\end{lemma}

\noindent{\bf Proof}

For $m>1\!+\!\sigma,\; m\in\Z^+,\; \sigma>0$, we have
$-(m\!-\!\sigma)<-1<0$ and $(m-\sigma-1)>0$.  The first condition,
$-(m-\sigma)<0$, allows us to use the equation (\ref{frac_int}) to write
$$ D_{\!x|a}^{-(m-\sigma)} f(x)\,=\, \frac{1}{\Gamma(m-\sigma)}\int_a^x\!\!
\frac{f(t)}{(x\!-\!t)^{1+\sigma-m}}\;dt\;. $$
From (\ref{e:Dxfm}),
\begin{eqnarray}
\lefteqn{D^m_{\!x|a} \left( D_{\!x|a}^{-(m-\sigma)} f(x) \right)}\nn\\
&=& \frac{d^m}{dx^m} \left( \frac{1}{\Gamma(m\!-\!\sigma)}\int_a^x\!\!
\frac{f(t)}{(x\!-\!t)^{1+\sigma-m}}\;dt \right)\nn\\
&=& \frac{1}{\Gamma(m\!-\!\sigma)}\int_a^x f(t) \,
\left( \frac{d^m}{dx^m} (x\!-\!t)^{m-\sigma-1} \right) dt\;.
\label{e:lemma_choice}
\end{eqnarray}
The second condition, $(m\!-\!\sigma\!-\!1)>0$, and the condition
$m>0$ allow us to use the second case of (\ref{e:Dxfm}). Thus, 
(\ref{e:lemma_choice}) becomes
\begin{eqnarray}
\lefteqn{\frac{1}{\Gamma(m\!-\!\sigma)}\int_a^x f(t) \, 
\frac{\Gamma(m\!-\!\sigma)}{\Gamma(-\sigma)} \, (x\!-\!t)^{-(1+\sigma)}
\;dt}\nn\\
&=&\frac{1}{\Gamma(-\sigma)}\int_a^x\!\!
\frac{f(t)}{(x\!-\!t)^{1+\sigma}}\;dt\nn\\
&=&D^\sigma_{\!x|a} f(x)\nn \;.
\end{eqnarray}
\hfill \qed

When $a=0$, (\ref{frac_int}) for $f(x) = x^r$ is well-defined only for
the half plane $r>-1$. Consequently, in the Riemann-Liouville
Fractional Derivative, $D^\sigma_{x|a} x^r$ is well-defined only for
the half plane $r>-1$.
\begin{equation}
D^\sigma_{\!x} x^r = \left\{\begin{array}{ll}
\ds \frac{\Gamma(1\!+\!r)}{\Gamma(1\!+\!r\!-\!\sigma)}\,
x^{r-\sigma} & (\sigma>0, \;r>-1)\\
\ds \bigg. x^r & (\sigma=0, \;\forall \; r)\\
\ds \Bigg. \frac{\Gamma(1\!+\!r)}{\Gamma(1\!+\!r\!-\!\sigma)}\,
\hat{x}^{r-\sigma}\Big|_a^x & (\sigma<0, \;r>-1)
\end{array}\right..
\label{RL_half_plane}
\end{equation}

\begin{definition}[{{\em Type II \ Fractional Derivative}}]\quad\\
  The ordinary derivative (derivative of integer order) of a constant
  is zero and ordinary derivatives commute. The fractional derivative
  is to inherit these properties from the ordinary derivative, i.e.
  fractional derivatives commute, and the fractional derivative of a
  constant is zero.
\end{definition}

For instance, given that $c$ is an arbitrary constant and
$D^1_{\!x}\,c\,=\,0$,
$$D^\sigma_{\!x}\,c\,=\,D^{(\sigma-1)}_{\!x}\!
\left(D^1_{\!x}\,c\right)\,=\,D^{(\sigma-1)}_{\!x}\,0\,=\,0
\quad (\sigma>1).$$
Since $D^r_{\!x}x^r\,=\,\Gamma(1+r)\;\; (r\ge0)$ and
$D^\sigma_{\!x}\,c\,=\,0$, by continuity,
$$D^\sigma_{\!x} x^r\,=\,0\quad (\sigma>r,\;r\ge0)$$
where the commutativity of fractional derivatives is
preserved.  Hence, a {\em Type II} \ Fractional Derivative with
commuting fractional derivatives is
\begin{equation}
D^\sigma_{\!x} x^r = \left\{\begin{array}{ll}
\ds 0 & (\sigma>r,\;r\ge0)\\
\ds \Bigg. \frac{\Gamma(1\!+\!r)}{\Gamma(1\!+\!r\!-\!\sigma)}\, x^{r-\sigma} & 
(0<\sigma\le r, \;r\ge 0) \mbox{ and } (\sigma>0, \;-1<r<0)\\
\ds \bigg. x^r & (\sigma=0, \;\forall \; r)\\
\ds \Bigg. \frac{\Gamma(1\!+\!r)}{\Gamma(1\!+\!r\!-\!\sigma)}\,
\hat{x}^{r-\sigma}\Big|_a^x & (\sigma<0, \;r>-1)
\end{array}\right..
\end{equation}

\medskip

Note that there is a mutual trade-off between the properties of
analyticity and commutativity in these two types:

\smallskip

\noindent {\em Type I}$\,$: we choose to preserve {\em global analyticity} and
lose {\em global commutativity}.

\smallskip

\noindent {\em Type II}$\,$: we choose to preserve {\em global commutativity}
and lose {\em global analyticity}.

\smallskip

\subsection{An Application of {\em Type II}\,: Power Series of Non-integer 
Order}

For {\em Type I}$\;$, we write the power series for $\exp(x)$ as
\begin{equation}
\exp(x)\,=\,\lim_{\epsilon\to 0}\sum_{k=-\infty}^{\infty}
\frac{1}{\Gamma(1\!+\!\epsilon\!+\!k)}\,x^{k+\epsilon}\;,
\end{equation}
and
\begin{equation}
D^\sigma_{\!x} \,\lim_{\epsilon\to 0}\sum_{k=-\infty}^{\infty} 
\frac{1}{\Gamma(1\!+\!\epsilon\!+\!k)}\,x^k 
= \lim_{\epsilon\to 0}\sum_{k=-\infty}^{\infty} 
\frac{1}{\Gamma(1\!+\!\epsilon\!+\!k\!-\!\sigma)}\,x^{k-\sigma+\epsilon} 
\quad (\sigma>0)\;.
\end{equation}

\bigskip

For {\em Type II}$\,$, we write it as
\begin{equation}
\exp(x)\,=\,\sum_{k=0}^{\infty}\frac{1}{\Gamma(1\!+\!k)}\,x^k\;,
\label{e:exppowerseries}
\end{equation}
and
\begin{equation}
D^\sigma_{\!x} \sum_{k=0}^{\infty} \frac{1}{\Gamma(1\!+\!k)}\,x^k = 
\!\!\sum_{k\,=\left\lceil\sigma\right\rceil}^{\infty} 
\!\frac{1}{\Gamma(1\!+\!k\!-\!\sigma)}\,x^{k-\sigma} 
\quad (\sigma>0)
\end{equation}
where $\left\lceil\big.\;\;\right\rceil$ denotes taking the integer
ceiling, since $D^\sigma_{\!x} x^r = 0 \quad (\sigma>r,\;r\ge 0)$.

In general, for {\em Type II}$\,$,
\begin{equation}
D^\sigma_{\!x} \sum_{k=0}^{\infty} a_k x^k = 
\!\!\sum_{k\,=\left\lceil\sigma\right\rceil}^{\infty} 
\!a_k\,\frac{\Gamma(1\!+\!k)}{\Gamma(1\!+\!k\!-\!\sigma)}\,x^{k-\sigma} 
\quad (\sigma>0) \label{nonint_pow_series}
\end{equation}
where $\left\lceil\big.\;\;\right\rceil$ denotes taking the integer
ceiling.

When $\sigma$ is not an integer, the series on the right of
(\ref{nonint_pow_series}) forms a {\em power series of non-integer order}.

\begin{definition}[Notation]\quad\\
Define the notation
\begin{equation}
f(\sigma,x) \equiv D^\sigma_{\!x}f(x)\;.
\end{equation}
\end{definition}
Think of $\sigma$ in the following way: the one-variable function $f(x)$ is
extended to a two-variable function $f(\sigma,x)$ in which $\sigma$
has now become a variable of the extended function.

For instance,
\begin{eqnarray}
\lefteqn{\cos(\sigma,x)\,=\,D^\sigma_{\!x}\cos(x)
\,=\,\ds D^\sigma_{\!x}\sum_{k=0}^{\infty}
\frac{(-1)^k}{(2k)!}\,x^{2k}}\nn\\
&&=\sum_{k\,=\,W\!c(\sigma)}^{\infty}\!
\frac{(-1)^k}{(2k)!}\,\frac{\Gamma(1\!+\!2k)}
{\Gamma(1\!+\!2k\!-\!\sigma)}\,x^{2k-\sigma}\;\;
(\sigma>0)\\
&&\quad\mbox{where }\quad
W\!c(\sigma)=\left\lceil\frac{\sigma}{2}+1\right\rceil-1\;,\nn
\end{eqnarray}
\begin{eqnarray}
\lefteqn{\sin(\sigma,x)\,=\,D^\sigma_{\!x}
\sin(x)\,=\,\ds D^\sigma_{\!x}\sum_{k=0}^{\infty}
\frac{(-1)^k}{(2k\!+\!1)!}\,x^{2k+1}}\nn\\
&&=\ds\sum_{k\,=\,W\!s(\sigma)}^{\infty}\!
\frac{(-1)^k}{(2k\!+\!1)!}\,\frac{\Gamma(2(k\!+\!1))}
{\Gamma(2(k\!+\!1)\!-\!\sigma)}\,
x^{2k+1-\sigma}\;\;(\sigma>0)\\
&&\qquad\mbox{where }\quad
W\!s(\sigma)=\left\lceil\frac{\sigma}{2}+\frac{1}{2}\right\rceil-1\;,\nn
\end{eqnarray}
and
\begin{eqnarray}
\lefteqn{\exp(\sigma,x)\,=\,D^\sigma_{\!x}
\exp(x)\,=\,\ds D^\sigma_{\!x}\sum_{k=0}^{\infty}
\frac{(-1)^k}{k!}\,x^k}\nn\\
&&=\ds\sum_{k\,=\,\left\lceil\sigma\right
\rceil}^{\infty}\!\frac{(-1)^k}{k!}\,\frac{\Gamma(1\!+\!k)}
{\Gamma(1\!+\!k\!-\!\sigma)}\,
x^{k-\sigma}\;\;(\sigma>0)\;.
\end{eqnarray}

\begin{table}[hbt]
$$\begin{array}{|c||c|c|c|c|c|c|c|c|c|c|c|}
\hline \sigma         &0.0&0.5&1.0&1.5&2.0&2.5&3.0&3.5&4.0&4.5&\cdots\\
\hline W\!c(\sigma)&0  &1  &1  &1  &1  &2  &2  &2  &2  &3&\cdots\\
\hline W\!s(\sigma)&0  &0  &0  &1  &1  &1  &1  &2  &2  &2&\cdots\\
\hline \hline\end{array}$$
\caption{Some tabulated values of $W\!c(\sigma)$ and $W\!s(\sigma)$.}
\end{table}

Note that
$$\left.\begin{array}{r}\cos(\sigma,0)\\\sin(\sigma,0)\\
\exp(\sigma,0)\end{array}\right\}=\,1\;\;\mbox{if}\;\;\sigma\in
\left\{\begin{array}{l}2\,\Z\\2\,\Z\!+\!1\\
\Z\end{array}\right\},\;0\;\;\mbox{otherwise};$$
$$\begin{array}{rclcl}
\cos(\sigma,x)&=&\pm\,\sin(\sigma\!\pm\!1,x)&=&-\cos(\sigma\!\pm\!2,x)\;,\\
\sin(\sigma,x)&=&\mp\cos(\sigma\!\pm\!1,x)&=&-\,\sin(\sigma\!\pm\!2,x)\;,\\
\exp(\sigma,x)&=&\;\;\;\exp(\sigma\!\pm\!1,x)\;,\end{array}$$
in agreement with the definitions of $\,\cos(x),\;\sin(x)$ and
$\exp(x)\,$ when $\,\sigma\in\Z\,$.

\begin{figure}[hbt]
\begin{center}
\begin{tabular}{llll}
$\cos(0.01,\!x)$&$\cos(0.25,\!x)$&$\cos(0.5,\!x)$&$\cos(0.75\!,x)$\\
\includegraphics[height=80pt]{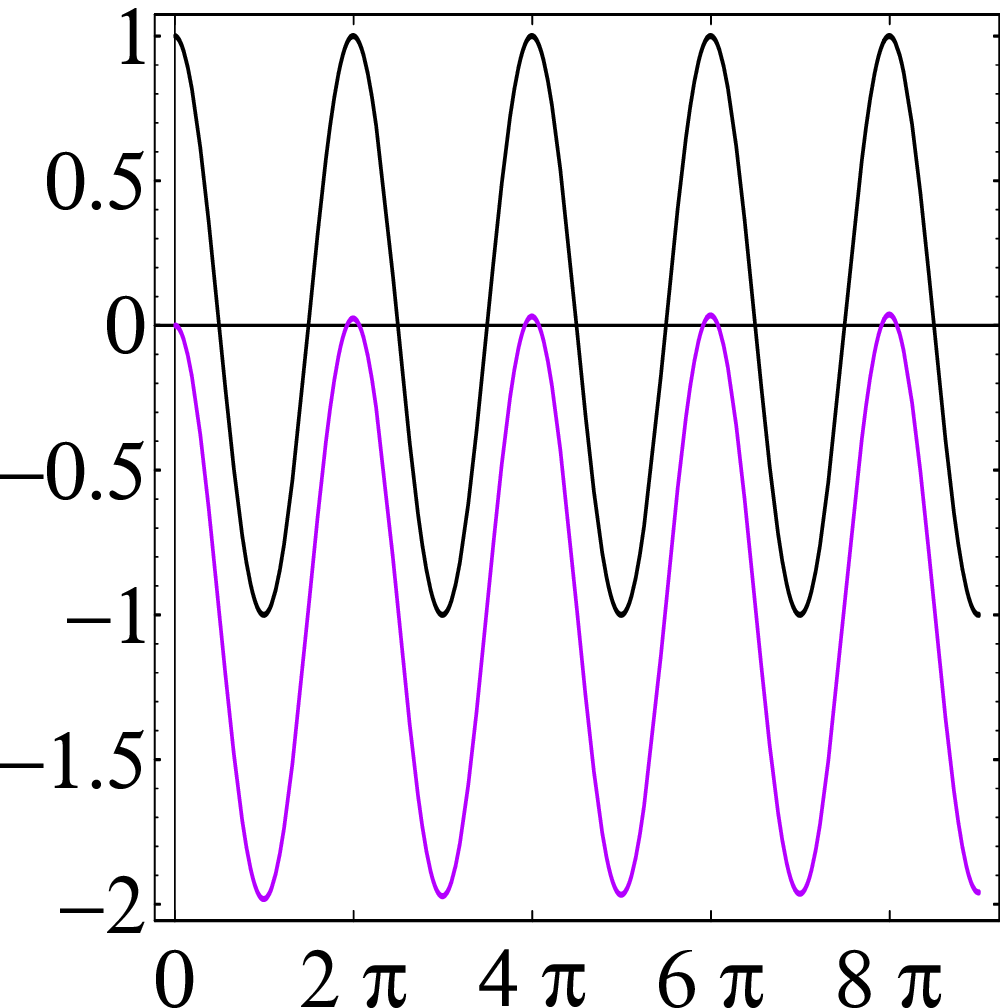}&
\includegraphics[height=80pt]{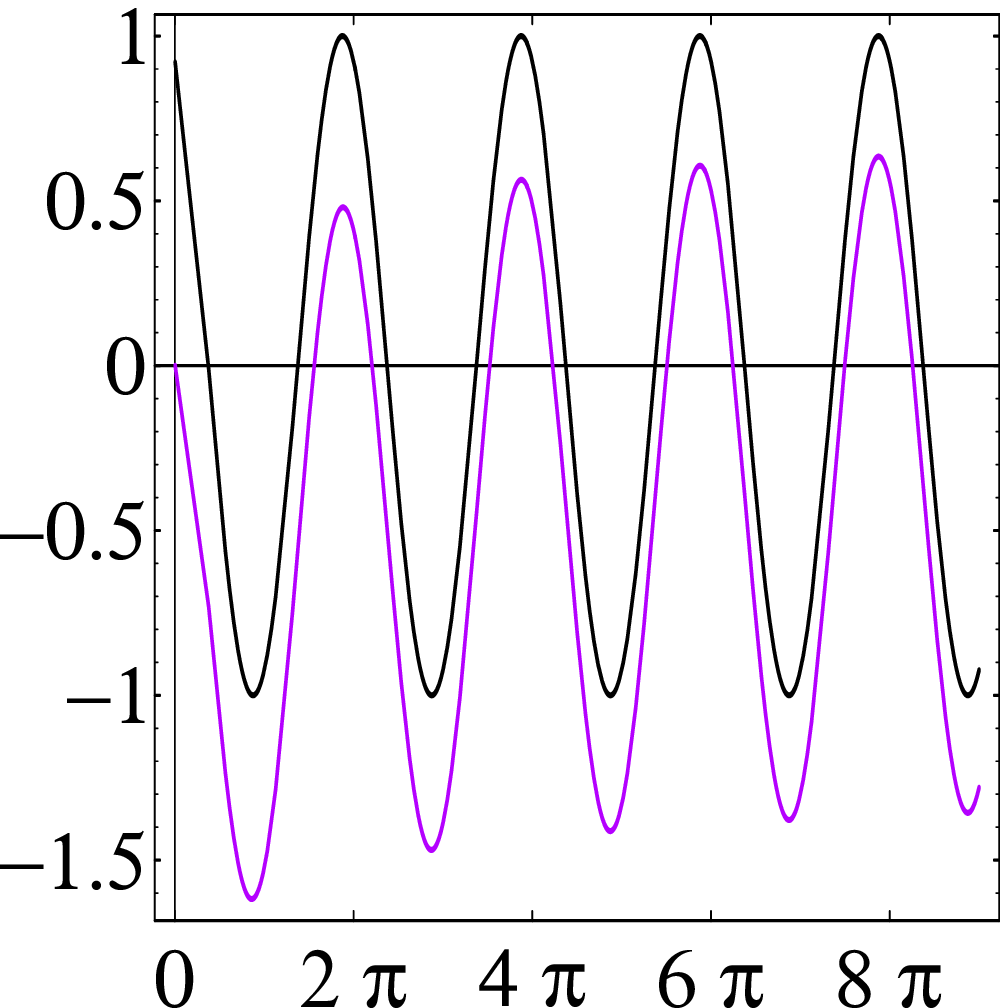}&
\includegraphics[height=80pt]{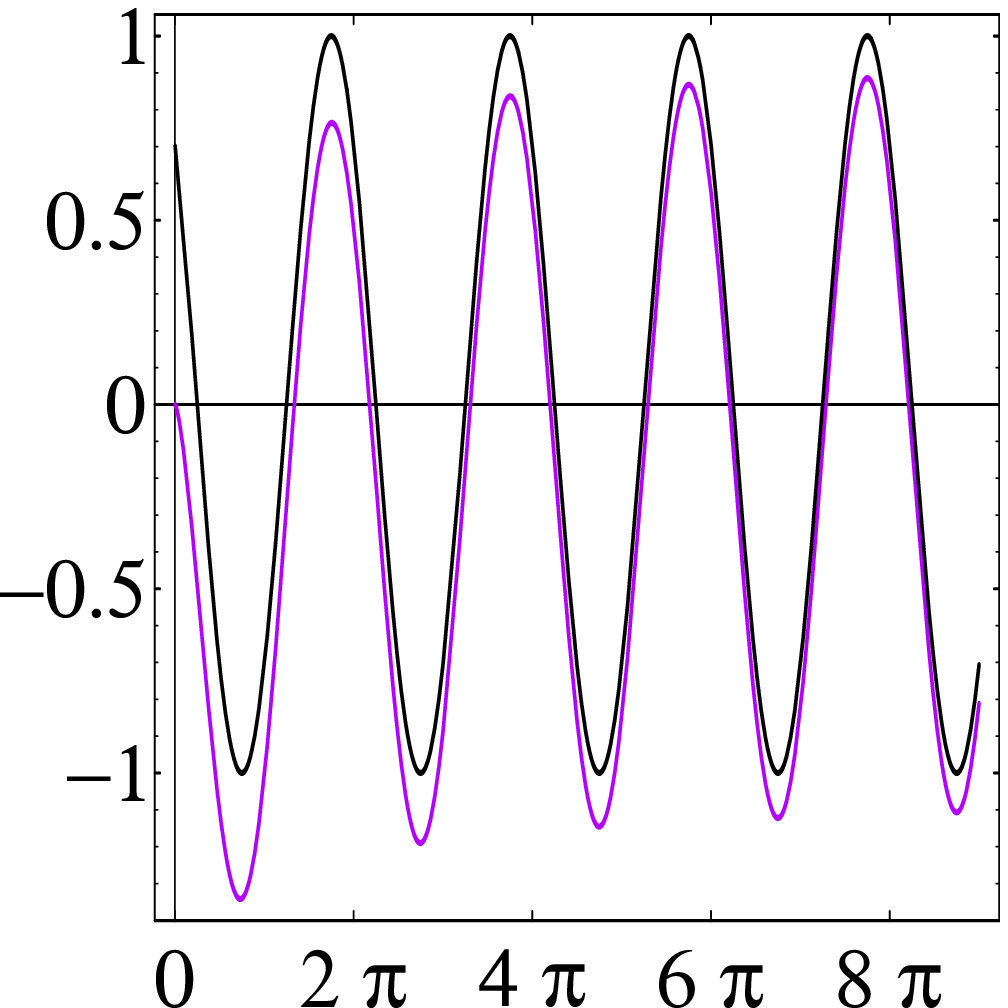}&
\includegraphics[height=80pt]{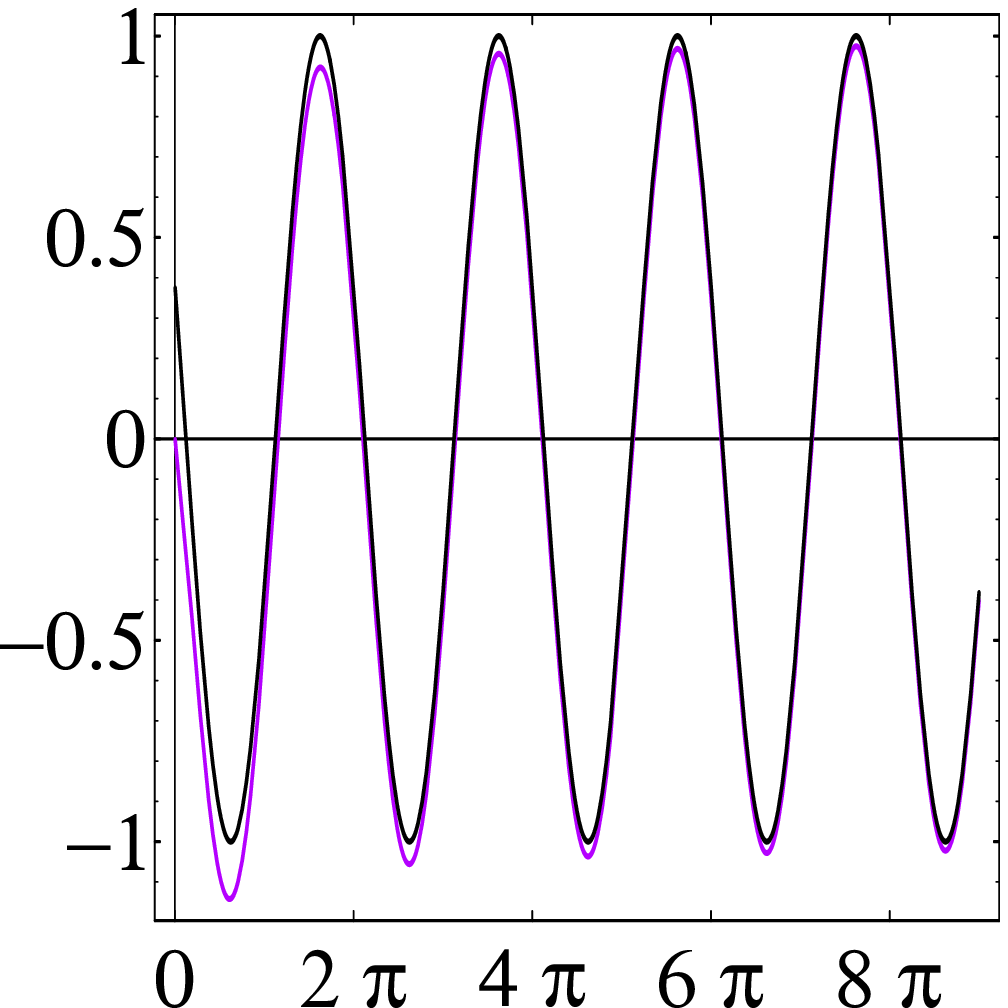}\\
$\qquad\qquad x$&$\qquad\qquad x$&$\qquad\qquad x$&$\qquad\qquad x$\\
\\
$\cos(1.25,\!x)$&$\cos(1.5,\!x)$&$\cos(1.99,\!x)$&$\cos(2.01,\!x)$\\
\includegraphics[height=80pt]{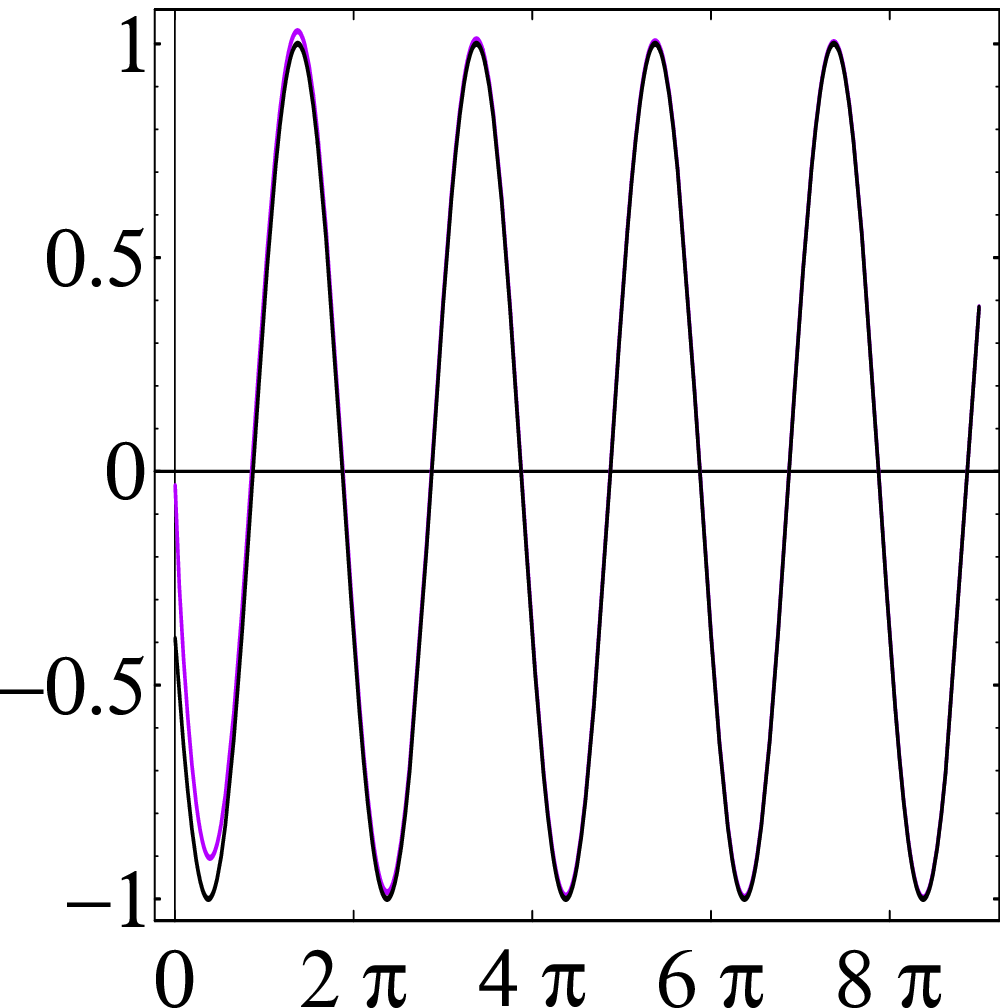}&
\includegraphics[height=80pt]{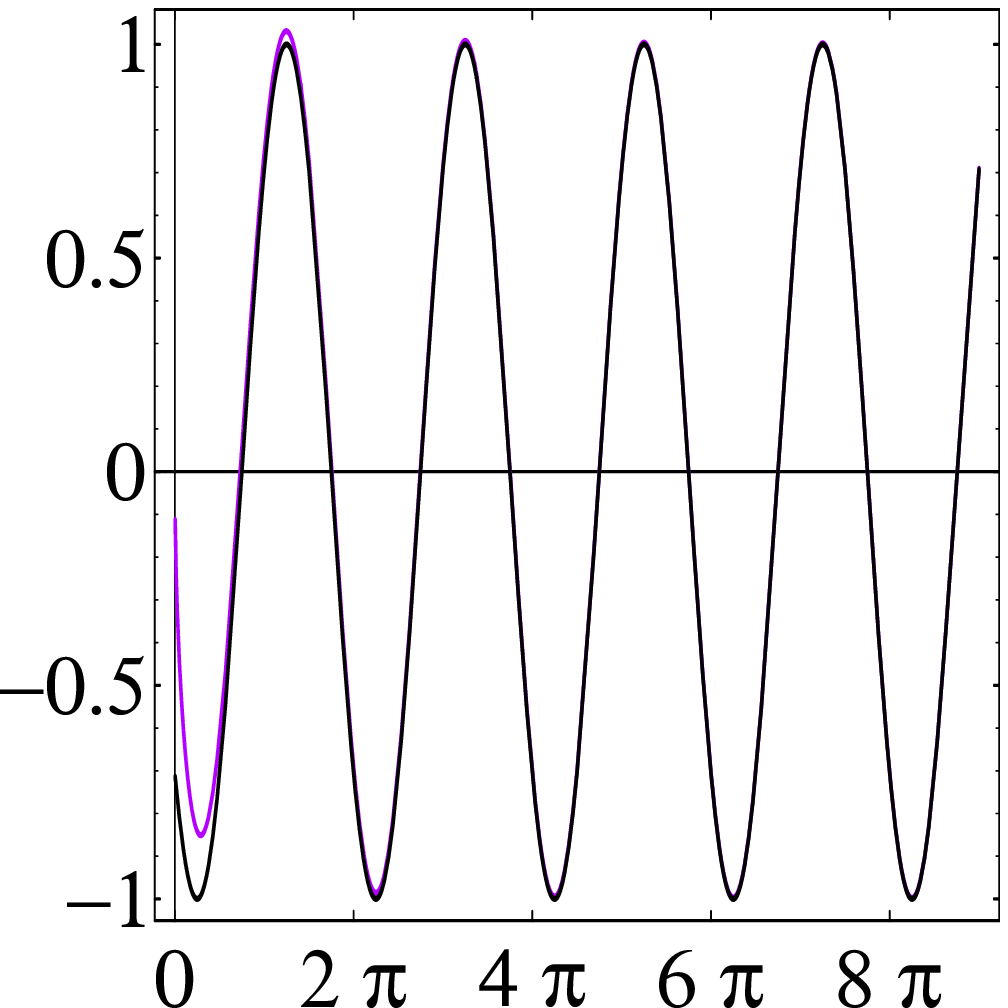}&
\includegraphics[height=80pt]{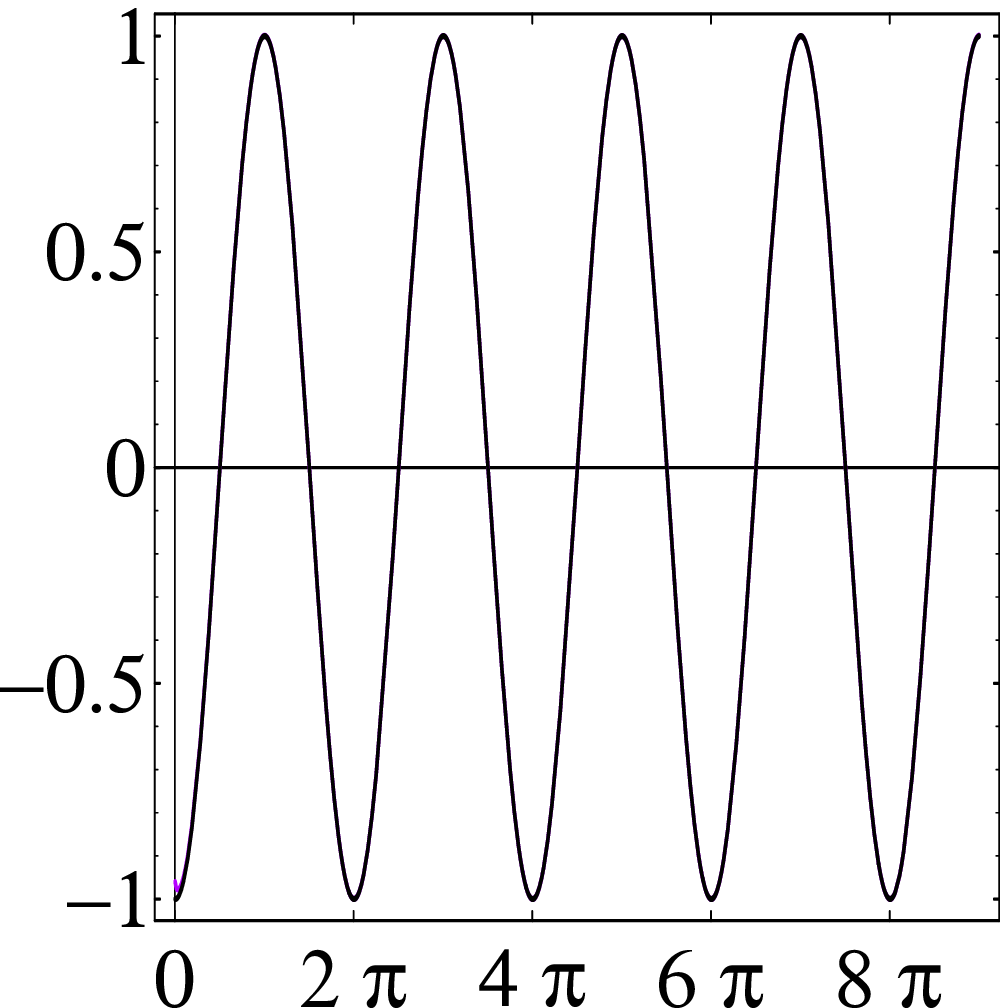}&
\includegraphics[height=80pt]{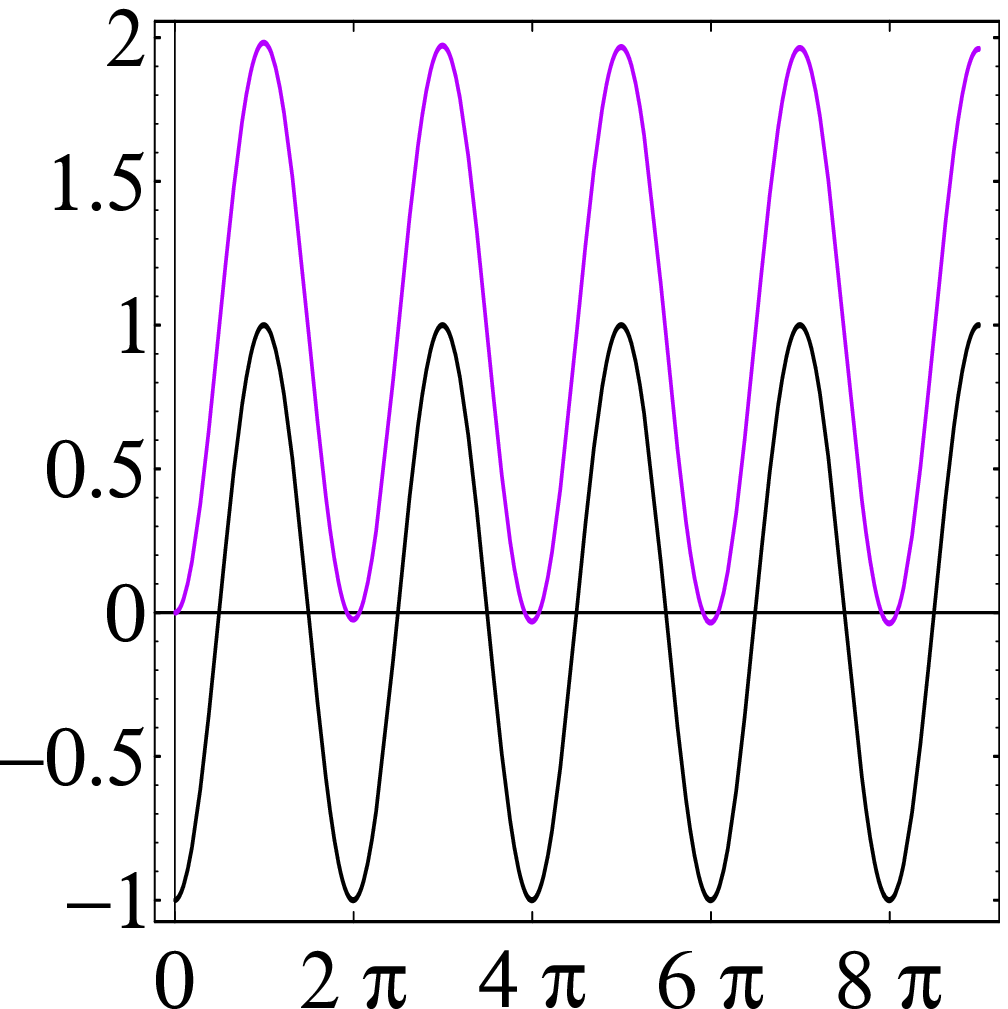}\\
$\qquad\qquad x$&$\qquad\qquad x$&$\qquad\qquad x$&$\qquad\qquad x$
\end{tabular}
\caption{Asymptotic limit of $\cos(\sigma,x)\equiv 
D^\sigma_{\!x}\cos(x)\sim\cos(x\!+\!\pi\sigma/2)$ as $x\to\infty$.}
\label{fig:cos_asymp_lim}
\end{center}
\end{figure}

\begin{figure}[hbt]
\begin{center}
\begin{tabular}{c}
$\left(\Big.\exp(\sigma,x) - \exp(x)\right)\qquad\qquad\qquad$\\
\includegraphics[height=140pt]{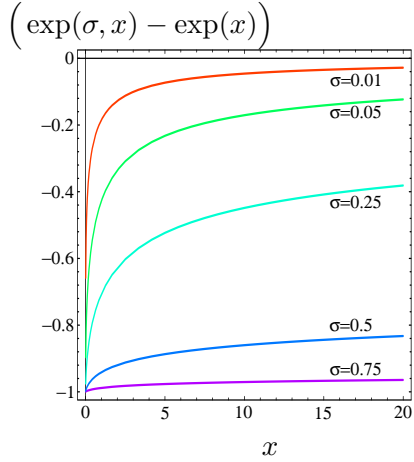}\\
$\qquad\quad x$
\end{tabular}
\caption{Asymptotic limit of $\exp(\sigma,x) - \exp(x) \to 0\;$ as 
$\,x\to\infty$.}
\label{fig:exp_asymp_lim}
\end{center}
\end{figure}

It can be observed from Figures \ref{fig:cos_asymp_lim} \ and
\ref{fig:exp_asymp_lim} \ that there exist asymptotic limits
\begin{equation}
\left.\begin{array}{rcl}
  \bigg.\cos(\sigma,x)&\!\!\sim\!&
  \cos(x\!+\!\ds\frac{\pi}{2}\,\sigma)\\
  \bigg.\sin(\sigma,x)&\!\!\sim\!&\,
  \sin(x\!+\!\ds\frac{\pi}{2}\,\sigma)\\
  \big.\exp(\sigma,x)&\!\!\sim\!&
  \exp(x)
\end{array}\right\}\;\forall\;\sigma\;\mbox{ as }\;x\to\infty\;.
\end{equation}

In general, any function in a power series representation can
similarly have a power series of non-integer order generalization.

\vfill

\subsection{Concrete Representations of Pseudo-Groups}

Consider extending the group elements of $SO(2)$ (the group of
rotation in a plane) as follows:

\begin{eqnarray}
&&R(\theta) = \left( \begin{array}{rl}
 \cos\,\theta & \sin\,\theta\\
-\sin\,\theta & \cos\,\theta
\end{array}\right)\;\;\mbox{where }\theta\in [0,2\,\pi)\nn\\
&&\quad\longmapsto
\begin{array}{c}R(\sigma,\theta) = \left( \begin{array}{rl}
 \cos(\sigma,\theta) & \sin(\sigma,\theta)\\
-\sin(\sigma,\theta) & \cos(\sigma,\theta)
\end{array}\right)\;\;\mbox{where }\theta\in[0,\infty)
\end{array}\!\!.
\end{eqnarray}
$R(\sigma,\theta)$ forms a family of sets, parametrized at 2 levels. The
family of sets is parametrized by $\sigma$, and each of these sets is
further parametrized by $\theta$. Denote the family of these sets as
$SO(2;\sigma,\theta)$.

Since $R(0,\theta)\in SO(2)\;\forall\,\theta\,$ and
$R(\sigma,\theta)\sim R(0,\theta+\pi\sigma/2)$ as $\theta\to\infty$,
we are motivated to introduce the idea of pseudo-groups as in
Definition \ref{d:pseudo-group}.

\begin{figure}[hbt]
\begin{center}
\includegraphics{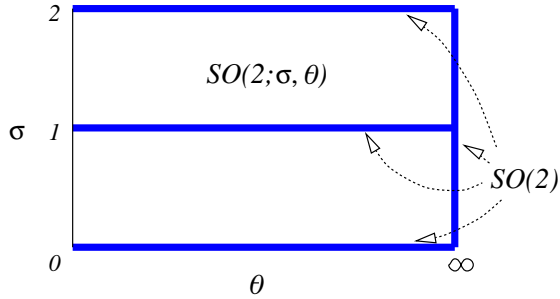}
\caption{$SO(2;\sigma,\theta)$ plane diagram.}
\label{fig:SO2}
\end{center}
\end{figure}

$SO(2;\sigma,\theta)$ is a pseudo-group of $SO(2)$ since it is
isomorphic to $SO(2)$ \\
either when the parameter ${\sigma\to n\in\Z}$ while
$\,\theta\,$ varies freely in the interval $[\,0,\infty)$,
\begin{eqnarray}
  &&R(\sigma,\theta)\;R(\sigma',\theta)\,\sim\,
  R(\sigma+\sigma',\theta)\;\;\mbox{as}\;\;\sigma,\sigma'\!\to
  n,n'\!\in\Z\nn\\
  &&\quad\Rightarrow\;\;\lim_{\sigma\to n\in\Z}
  SO(2;\sigma,\theta) \,\cong\,SO(2)
\end{eqnarray}
or when the parameter ${\theta\to\infty}$ while $\sigma$ varies freely
in the interval $(0,2)$,
\begin{eqnarray}
  &&R(\sigma,\theta)\;R(\sigma,\theta')\,\sim\,
  R(0,\theta+\ds\frac{\pi}{2}\sigma)\;
  R(0,\theta'\!+\ds\frac{\pi}{2}\sigma)\nn\\
  &&\quad=\,R(0,\theta+\theta'\!+\pi\sigma)
  \;\;\mbox{as}\;\;\theta,\theta'\to\infty\nn\\
  &&\quad\Rightarrow\;\lim_{\theta\to\infty} 
  SO(2;\sigma,\theta)\,\cong\,SO(2)
\end{eqnarray}
as shown in Figure \ref{fig:SO2}.

\section{Measure of Group Property Deviation}

\begin{definition}[A Measure of Group Property Deviation]\quad\\
\label{d:measuredeviation}
  We define a measure of group property deviation, ${\cal
    W}(G(\sigma,x),G\,\big|\,\sigma,x)$, for a pseudo-group $G(\sigma,x)$,
  as the measure of how much group property the pseudo-group has lost or
  deviated from the associated ``parent'' group $G$ from which it is
  analytically extended.  When the pseudo-group becomes isomorphic to
  the parent group for certain values of the parameter, the measure
  should be zero.
\end{definition}

\begin{figure}[hbt]
\begin{center}
\includegraphics[height=180pt]{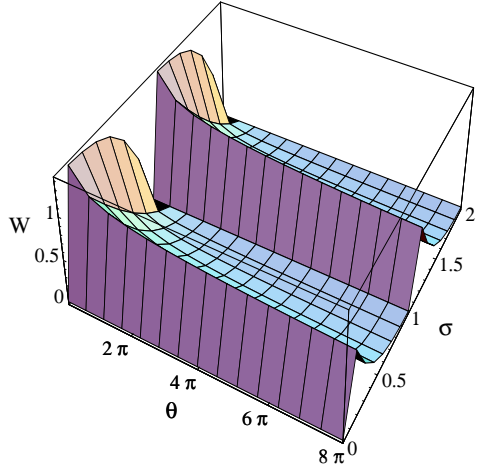}
\caption{Measure ${\cal W}$ of $SO(2;\sigma,\theta)$.}
\label{fig:SO2_measure}
\end{center}
\end{figure}

For the case of pseudo-group $SO(2;\sigma,\theta)$,
\begin{eqnarray}
\lefteqn{{\cal W}(SO(2;\sigma,\theta),SO(2)\,\big|\,\sigma,\theta)}\nn\\
&=&\left|\,\Big.R(\sigma,\theta) - R(\theta\!+\!\frac{\pi}{2}\sigma)\;
\right|\nn\\
&=&\left|\begin{array}{cc}\;\;\cos(\sigma,\theta)-\cos(\theta
\!+\!\ds\Bigg.\frac{\pi}{2}\sigma)&
\;\sin(\sigma,\theta)-\sin(\theta
\!+\!\ds\frac{\pi}{2}\sigma)\nn\\
-\Big(\sin(\sigma,\theta)-\sin(\theta
\!+\!\ds\Bigg.\frac{\pi}{2}\sigma)\Big)&
\;\cos(\sigma,\theta)-\cos(\theta
\!+\!\ds\frac{\pi}{2}\sigma)\end{array}\right|\nn\\
&=&\left|\begin{array}{cc}\;\;\delta\!\cos&\;\;\delta\!\sin\\
-\delta\!\sin&\;\;\delta\!\cos\end{array}\right|\nn\\
&=&\Bigg.\sqrt{\left(\delta\!\cos+\delta\!\sin\right)^2+
\left(-\delta\!\sin+\delta\!\cos\right)^2}\nn\\
&=&\Bigg.\sqrt{2\left(\delta\!\cos^{\,2}+\delta\!\sin^2\right)}\nn\\
&=&\ds\sqrt{2\left(\left(\cos(\sigma,\theta)-\cos(\theta
\!+\!\frac{\pi}{2}\sigma)\right)^{\!2}
+\left(\sin(\sigma,\theta)-\sin(\theta\!+\!
\frac{\pi}{2}\sigma)\right)^{\!2}\right)}
\end{eqnarray}
satisfies the requirement in Definition \ref{d:measuredeviation}. 
See Figure \ref{fig:SO2_measure} \ for the plot
of this measure.

Similarly for the case of $U(1;\sigma,i x)$, a pseudo-group of
$U(1)$ where $x\in\R$,
\begin{eqnarray*}
&&\exp(\sigma,i x_1)\;\exp(\sigma',i x_2)\,\sim\,
\exp(\sigma,i(x_1\!+\!x_2))
\quad\mbox{as}\;\;\sigma,\sigma'\to n,n'\in\Z\\
&&\quad\Rightarrow\quad\lim_{\sigma\to n\in\Z} U(1;\sigma,x)\,\cong
\,U(1)\;,\\
&&\exp(\sigma,i x_1)\;\exp(\sigma,i x_2 )\,\sim\,
\exp(\sigma,i(x_1\!+\!x_2))\quad\mbox{as}\;\;x_1 ,x_2 \to \infty\\
&&\quad\Rightarrow\quad\lim_{\theta\to\infty} U(1;\sigma,x)\,\cong
\,U(1)\;,\\
&&{\cal W}(U(1;\sigma,ix),U(1)\,\big|\,\sigma,i x)\,=\,
\exp(\sigma,i x)-\exp(i x)\;.
\end{eqnarray*}
This measure was plotted in Figure \ref{fig:exp_asymp_lim}.

\section{Rotations and Deformations of $SO(2;\sigma,\theta)$}

\begin{figure}[hbt]
\begin{center}
$\begin{array}{l}\;\;\fbox{$\sigma\Big\backslash\theta\;$}\;\,\approx 
0\;\,\pi/16\;\;\pi/8\;\;\;\;\pi/4\;\;\;\pi/2\;\;\,3\pi/4\;\;\;\,\pi
\;\;\;\;3\pi/2\;\;\;\, 2\pi\quad\; 4\pi\quad\;\, 6\pi \quad\;\, 8\pi\\
\begin{array}{rl}
\raisebox{168pt}{$\begin{array}{r}0\\\\0.001\\\\0.25\\\\0.5\\\\0.75
\\\\1\\\\1.25\\\\1.5\\\\1.75\\\\2\\\\3\\\\4 \end{array}$}
&\bspace
\includegraphics[height=341pt]{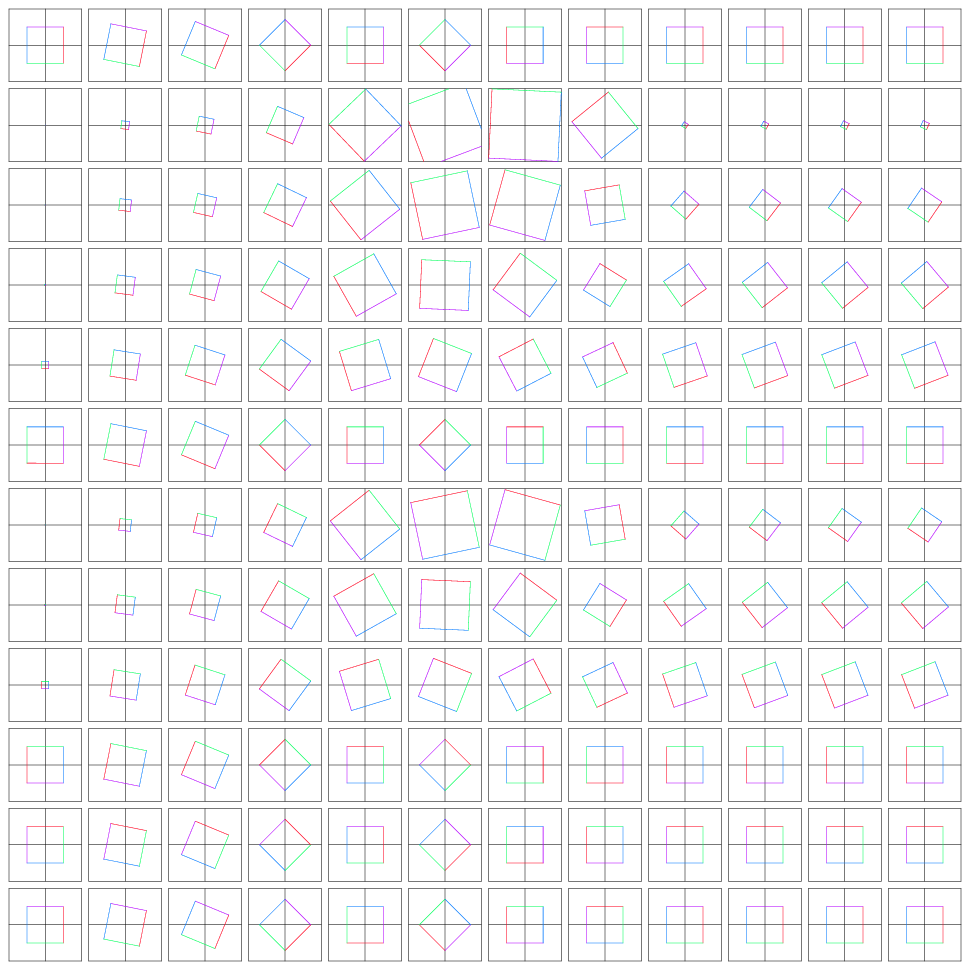}
\end{array}\end{array}$
\caption{$SO(2;\sigma,\theta)$ rotations and deformations}
\label{fig:SO2_rot}
\end{center}
\end{figure}

Figure \ref{fig:SO2_rot} \ shows the effect of planar rotations and
deformations of $SO(2;\sigma,\theta)$ on a square with vertices
$\{\,(1,-1),\,(1,1),\,(-1,1),\,(-1,-1)\}$ in a sequence of $(x,y)$
planes clipped by square windows of size
$x\!\in\![-2,2],\;y\!\in\![-2,2]$. The deformation effects can be seen
as a combination of rotations and contractions/dilations.

\section{Symmetry Breaking/Deforming via Pseudo-Group}

In the Higgs mechanism of Spontaneous Symmetry Breaking \cite{higgs}
in Particle Physics and Cosmology,
\begin{enumerate}
\item the symmetry of the effective potential $V_{e\!f\!f}$ in a
  Lagrangian density $\cal L$ with respect to a gauge group $G$ is
  preserved, while
  
\item the symmetry of the Quantum state $\psi$ satisfying the
  equations of motion derived from $\cal L$ is broken and reduced from
  $G$ to a subgroup, $H\subset G$.
\end{enumerate}

The profile of $V_{e\!f\!f}$ changes with energy or temperature.  At
high energy or temperature, the symmetry of $\psi$ is restored from
$H\to G$.

In the case of symmetry breaking/deforming via pseudo-group, the
situation is very different from the Higgs mechanism.  The symmetry in
a group $G$ itself is broken to a subgroup $H\subset G$ or
``deformed'' into a pseudo-group with an approximate symmetry of $G$.

Consider $SO(3;(\sigma_1,x_1),(\sigma_2,x_2))$, a pseudo-group of
$SO(3)$, as an example.

When both $\sigma_1,\sigma_2=0$, $SO(3;(\sigma_1,x_1),(\sigma_2,x_2))
\sim SO(3)$.

However, when $\sigma_1\not\in\Z$ and $\sigma_2=0$, the $SO(3)$
symmetry is broken, approximate or restored, for small, intermediate,
or large values of $x$ respectively.  The $SO(3)$ symmetry in a sphere
is ``deformed'' to an approximate $SO(3)$ symmetry or completely
broken to $SO(2)$ in a plane depending on the chosen values of
$\sigma_1$ and $x$.

When both $\sigma_1,\sigma_2\not\in\Z$, the $SO(2)$ is further
``deformed'' to an approximate $SO(2)$ symmetry or broken to identity.

The symmetry breaking/deforming sequence is thus
\begin{eqnarray}
SO(3;(\sigma_1,x_1),(\sigma_2,x_2))
\stackrel{\sigma_1,\,\sigma_2=\,0}{\to\!\!\!-\!\!\!-\!\!\!\longrightarrow}
SO(3)\nn\\
\stackrel{\sigma_1\not\in\,\Z\,,\,\sigma_2=\,0}
{\hookrightarrow\!\!\!-\!\!\!-\!\!\!-\!\!\!-\!\!\!-\!\!\!\longrightarrow}
SO(2) \stackrel{\sigma_1,\,\sigma_2\not\in\,\Z}
{\to\!\!\!-\!\!\!-\!\!\!\longrightarrow}\,\id\;.
\end{eqnarray}

Similarly for $SU(N;(\sigma_1,x_1),(\sigma_2,x_2),\cdots,(\sigma_N,x_N))$,
the symmetry breaking/deforming sequence is
$$\begin{array}{l}
SU(N;(\sigma_1,x_1),(\sigma_2,x_2),\cdots,(\sigma_N,x_N))\\ \\
\begin{array}{l}
\quad\stackrel{\sigma_1,\,\sigma_2,\cdots,\,\sigma_N=\,0}
{\hookrightarrow\!\!\!-\!\!\!-\!\!\!-\!\!\!-\!\!\!-\!\!\!-\!\!\!-\!\!\!
-\!\!\!-\!\!\!-\!\!\!-\!\!\!-\!\!\!-\!\!\!-\!\!\!-\!\!\!-\!\!\!-\!\!\!
\longrightarrow}\quad SU(N)\\
\qquad\stackrel{\sigma_1\not\in\,\Z\,,\,\sigma_2,\cdots,\,\sigma_N=\,0}
{\hookrightarrow\!\!\!-\!\!\!-\!\!\!-\!\!\!-\!\!\!-\!\!\!-\!\!\!-\!\!\!
-\!\!\!-\!\!\!-\!\!\!-\!\!\!-\!\!\!-\!\!\!-\!\!\!-\!\!\!-\!\!\!-\!\!\!
\longrightarrow}\quad SU(N\!-\!1)\\
\qquad\qquad\qquad\quad\vdots\qquad\vdots\\
\qquad\qquad\stackrel{\sigma_1,\cdots,\,\sigma_{N-2}\not\in\,\Z\,,\,
\sigma_{N-1},\,\sigma_N=\,0}
{\hookrightarrow\!\!\!-\!\!\!-\!\!\!-\!\!\!-\!\!\!-\!\!\!-\!\!\!-\!\!\!
-\!\!\!-\!\!\!-\!\!\!-\!\!\!-\!\!\!-\!\!\!-\!\!\!-\!\!\!-\!\!\!-\!\!\!
\longrightarrow}\quad SU(2)\\
\qquad\qquad\quad\stackrel{\sigma_1,\cdots,\,\sigma_{N-1}\not\in\,\Z\,,
\,\sigma_N=\,0}
{\hookrightarrow\!\!\!-\!\!\!-\!\!\!-\!\!\!-\!\!\!-\!\!\!-\!\!\!-\!\!\!
-\!\!\!-\!\!\!-\!\!\!-\!\!\!-\!\!\!-\!\!\!-\!\!\!-\!\!\!-\!\!\!-\!\!\!
\longrightarrow}\quad U(1)\\
\qquad\qquad\qquad\stackrel{\sigma_1,\,\sigma_2,\cdots,\,
\sigma_N\not\in\,\Z}
{\hookrightarrow\!\!\!-\!\!\!-\!\!\!-\!\!\!-\!\!\!-\!\!\!-\!\!\!-\!\!\!
-\!\!\!-\!\!\!-\!\!\!-\!\!\!-\!\!\!-\!\!\!-\!\!\!-\!\!\!-\!\!\!-\!\!\!
\longrightarrow}\quad\id\;.
\end{array}
\end{array}$$

This mode of symmetry breaking/deforming seems to have useful
implications for models in Particle Physics and Cosmology.  The
challenge is to develop the full theory of symmetry breaking/deforming
via pseudo-group to model the approximate symmetry or gauge groups in
the Universe we live in today.

In Particle Physics, the symmetry of the flavor of quarks are not
exact symmetry but only approximate symmetry of Gell-Mann's Eightfold
way $SU(3)$ \cite{SU(3)} or Georgi-Sheldon's GUT $\,SU(5)$
\cite{SU(5)} because quarks of different flavors have different
masses, and light quarks do not transform in the mixing matrix exactly
into heavy quarks.  The effects from the presence of massive gluons
interactions and glueballs also perturbs the symmetry away from an
exact symmetry.  It is possible that $SU(3)$ and $SU(5)$ may be
``deformed'' into pseudo-groups with approximate symmetries that will
fit the phenomenological data better.

In the context of Cosmology, consider taking a pseudo-group to
describe the product of the residual exact and approximate symmetries
in the present day Universe.  Let's set the pseudo-group variable
$x\propto T$, the temperature of the Universe.  As we go back in time,
the temperature $T$ goes up, $x$ goes up, and we find that the
approximate and other broken symmetries are gradually being restored.
The rate at which the symmetries are being restored will be dependent 
on the values of $\sigma_1,\sigma_2,\cdots,\sigma_N$, the parameters
of the pseudo-group. The fully restored symmetry will be the symmetry
of the parent group of the pseudo-group.  Qualitatively, this model
resembles the unification of gauge groups in Cosmology.

\section{Towards a General Theory}

\begin{definition}[{\bf General Definition of Pseudo-Group and Measure
  ${\cal W}$}]\quad\\
An element of any Lie Group $G$ can be expressed as
$\exp(g.\,x)$ where $g$ is the generator of $G$ and $x$ is the
parameter of $G$.

The element of the pseudo-group of $G$ is thus
\begin{eqnarray}
\lefteqn{\exp(\sigma,g.\,x)\,=\,D^\sigma_{\!x}
\exp(g.\,x)\,=\,\ds D^\sigma_{\!x}\sum_{k=0}^{\infty}
\frac{(-1)^k}{k!}\,{(g.\,x)}^k}\nn\\
&&=\ds\sum_{k\,=\,\left\lceil\sigma\right
\rceil}^{\infty}\!\frac{(-1)^k}{k!}\,\frac{\Gamma(1\!+\!k)}
{\Gamma(1\!+\!k\!-\!\sigma)}\,
{(g.\,x)}^{k-\sigma}\;\;(\sigma\in\R,\sigma>0)\;.
\end{eqnarray}
\end{definition}

In the matrix representation of the generator $g$, $(g.\,x)$ is a
matrix and ${(g.\,x)}^{k-\sigma}$ is an analytic extension of the
matrix $(g.x)$.  We shall study the analytic extension of matrices and
outline the general steps in evaluating analytic extended matrices in
the next paper.

\bigskip

\noindent{\Large \bf Appendix}

\appendix

\section{Extension of {\em Type II} \ from 
$\;\sigma,r\in\R\;$ to $\;s,w\in\C$}

\begin{figure}[hbt]
\begin{center}
\includegraphics[height=120pt]{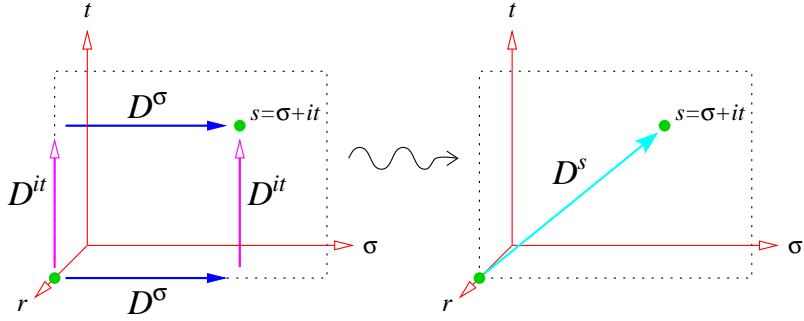}
\caption{Commutative arrows in $(s,r)$ diagram
of $D^s_{\!x} x^r$ in  {\em Type II}$\,$.}
\label{fig:D_commutative_diagram}
\end{center}
\end{figure}

The differential operator $d/dx$ commutes with itself and with its
inverse, the integral operator $\int dx$.  For {\em Type II}$\;$,
commutativity is preserved.  The {\em Type II}$\;$ commutative
operator $D^s, \;s\in\C$, acting on a function $f$ can be split into two
parts with each acting on the function independently,
\begin{equation}
D^s f = D^{(\sigma+it)} f = D^\sigma D^{it} f = D^{it} D^\sigma f \quad
(s=\sigma\!+\!it,\;\sigma,t\in\R)
\end{equation} 
as illustrated in the commutative diagram in Figure
\ref{fig:D_commutative_diagram}, with all the limits, if any, taken
only after the actions of all the operators have been performed.

Consider $D^s_{\!x} x^w, \;s=\sigma\!+\!it, \;w=u\!+\!iv$.\\

For $\sigma>u, \;u\ge 0$,
\begin{equation}
D^\sigma_{\!x} x^u = 0 \;\quad\Rightarrow\quad D^{it}_{\!x}
D^\sigma_{\!x} x^u = D^s_{\!x} x^u = 0\;.\label{e:Dzero1}
\end{equation}
Similarly, for $t>v, \;v\ge 0$,
\begin{equation}
D^{it}_{\!x} x^{iv} = 0 \quad\Rightarrow\quad D^\sigma_{\!x}
D^{it}_{\!x} x^{iv} = D^s_{\!x} x^{iv} = 0\;.\label{e:Dzero2}
\end{equation}

\begin{figure}[hbt]
\begin{center}
\includegraphics[height=125pt]{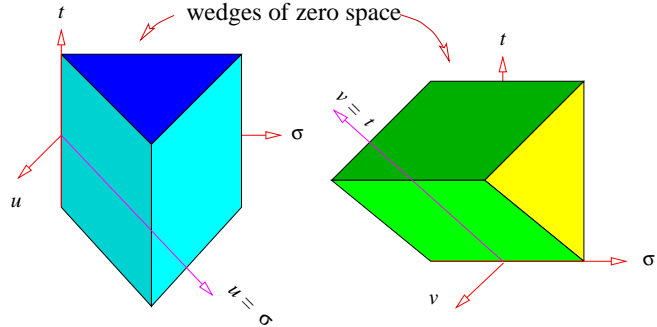}
\caption{The zero space wedges in $(s,w)$ diagrams where $D^s_{\!x} x^w$ 
in {\em Type II}$\,$.}
\label{fig:rszero}
\end{center}
\end{figure}

We can think of the following pictures. $D^{it}$ in (\ref{e:Dzero1})
expanding the triangular zero region of the $(\sigma,u)$ plane into a
wedge-shaped volume of infinite length along the $t$-direction.
Similarly $D^{\sigma}$ in (\ref{e:Dzero2}) expands the triangular zero
region of the $(it,iv)$ plane into another wedge-shaped volume of
infinite length along the $\sigma$-direction as shown in Figure
\ref{fig:rszero}. Both these wedge-shaped volume defines the zero
space in which $D^s_{\!x} x^w = 0$.

\section{Extension of Groups -- $\R_{\mbox{\rm (mod {\it n})}}$ Groups}
\label{sec:group}

The differential operator, and its inverse --- integral operator, can
act on different sets to generate different {\em discrete} groups.
These are groups of operators, ie. groups with operators as elements.

\begin{table}[hbt]
$$\fbox{$\ds\quad \frac{d^n}{dx^n} f(x) = f(x) \quad $}$$
$$\begin{array}{|l|l|l|}
\hline
\mbox{Order}&\quad\;f(x)&\qquad\quad\;\mbox{Symmetry Group}\\
\hline \hline
\,n\!=\!1 & \;\exp(x) & \ds \Z_1  = \left\{ \id
\right\}\!, \;\frac{d}{dx}\!\equiv\!\id\!=\!\frac{d^0}{dx^0}\\
\,n\!=\!2 & \left\{ \begin{array}{c}\!\!\cosh(x)\\
    \!\!\sinh(x)\end{array} \right.\!\!
& \ds \Z_2 = \left\{\id,\frac{d}{dx}\right\}\!, 
  \;\frac{d^2}{dx^2}\!\equiv\!\id\\
\,n\!=\!4 & \left\{ \begin{array}{c}\!\!\pm\cos(x)\\
    \!\!\pm\sin(x)\end{array} \right.\!\!
& \ds \Z_4 = \left\{\id,\frac{d}{dx},
  \frac{d^2}{dx^2},\frac{d^3}{dx^3}\right\}\!,
  \frac{d^4}{dx^4}\!\equiv\!\id\\
\hline \end{array} $$
\caption{Groups of $\ds\frac{d}{dx}$ acting on exponential, trigonometric
  and hyperbolic functions.}
\end{table}

\begin{figure}[hbt]
\begin{center}
\includegraphics{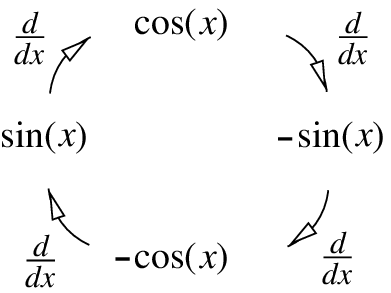}
\caption{$\Z_4$ group flow diagram of}
\begin{tabular}{c}
$\ds\frac{d}{dx}$ acting on set $\left\{\pm\cos(x), \pm\sin(x)\right\}$.
\end{tabular}
\label{fig:gpcos}
\end{center}
\end{figure}

Figure \ref{fig:gpcos} \ shows the $\Z_4$ group flow.

\begin{figure}[hbt]
\begin{center}
\includegraphics{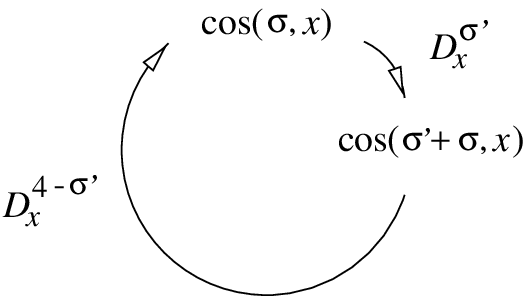}
\caption{$\R_{\mbox{(mod 4)}}$ group flow diagram of}
\begin{tabular}{c}
$D^{\sigma'}_{\!x}$ acting on set $\left\{
\cos(\sigma,x) \;\Big|\; \sigma\in[0,4)\right\}$.
\end{tabular}
\label{fig:gpcosigma}
\end{center}
\end{figure}

If elements of the set are extended from functions $f(x)$
to their analytic extensions $f(\sigma,x) = D^\sigma f(x)$ with
real $\sigma$, operators $D^{\sigma'}$ acting on these extended
sets will generate {\em continuous} groups or {\em Lie}
groups, eg., $D^{\sigma'}_{\!x}$ acting on the set 
$$\left\{ \cos(\sigma,x) \;\Big|\; \sigma\in[0,4), \;x\in[0,\infty)
\right\}$$
generates a natural analytic extension of the $\mbox{\bf Z}_4$ group,
$$\left\{ D^{\sigma'}_{\!x} \;\Big|\; \sigma'\in[0,4) \right\},
\;D^4_{\!x} \equiv \id$$
as illustrated in Figure \ref{fig:gpcosigma}.

By analogy to the concept of (mod $n$) congruence in Number Theory, we
denote this analytically extended group $\R_{\mbox{(\footnotesize mod 4)}}$.

\begin{table}[hbt]
$$\fbox{$\quad D^{\sigma'}_{\!x} f(\sigma,x) = f(\sigma,x) \quad $}$$
$$\!\!\begin{array}{|l|l|l|l|}
\hline
\mbox{Order}&\quad\;f(x)&\qquad\quad\;\mbox{Symmetry Group}\\
\hline \hline
\sigma'\!=\!1 &\;\exp(\sigma,x)&\R_{\mbox{(\footnotesize mod 1)}} 
  \!=\!\left\{ D^{\sigma'}_{\!x} 
\Big|\sigma' \in [0,1) \right\}\!, D^1_{\!x}\!\equiv\!\id \\
\sigma'\!=\!2 & \left\{ \begin{array}{c}\!\!\cosh(\sigma,x)\\
  \!\!\sinh(\sigma,x)\end{array} \right.\!\! & {\sf
R}_{\mbox{(\footnotesize mod 2)}}
  \!=\!\left\{ D^{\sigma'}_{\!x}\Big|\sigma'\in[0,2)\right\}\!,
  D^2_{\!x}\!\equiv\!\id \\
\sigma'\!=\!4 & \left\{ \begin{array}{c}\!\!\pm \cos(\sigma,x)\\
\!\!\pm\sin(\sigma,x)\end{array} \right.\!\! & \R_{
\mbox{(\footnotesize mod 4)}}
  \!=\!\left\{ D^{\sigma'}_{\!x}\Big|\sigma'\in[0,4)\right\}\!,
  D^4_{\!x}\!\equiv\!\id \\
\hline \end{array} $$
\caption{Analytically extended groups of $D^{\sigma'}$ acting on
  analytically extended exponential, trigonometric and hyperbolic
  functions.}
\end{table}

For complex $s=\sigma\!+\!it\,,\,s'=\sigma'\!+\!it'$,
$D^{s'}$ acting on set
$$\left\{\cos(s,x) \;\Big|\; \sigma \in [0,4) \,,\, t\in\R \,,\, x
  \in [0,\infty) \right\}$$
generates a Lie group $\,\R_{\mbox{(\footnotesize mod
    4)}}\,{\sf X}\,\R\,$ since $D^{it'}$ commutes with
$D^{\sigma'}$ and so $D^{it'}$ acts independently from $D^{\sigma'}$.

\begin{figure}[hbt]
\begin{center}
\includegraphics[height=110pt]{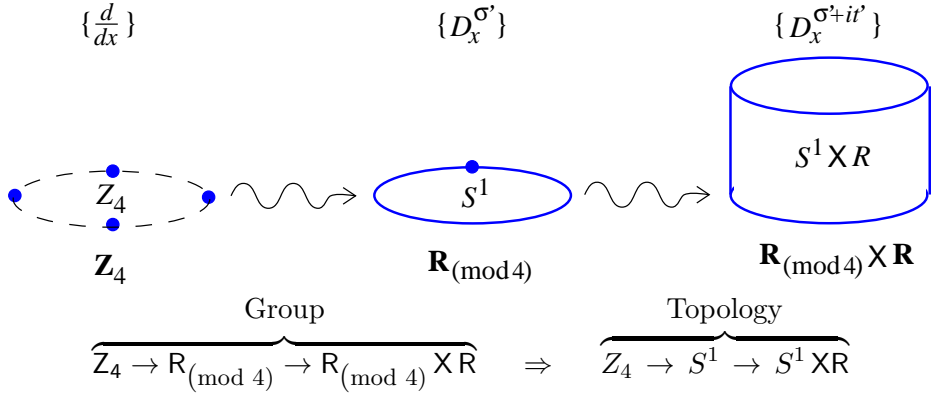}
\begin{tabular}{ccc}
Group & & Topology\\
$\overbrace{\Z_4 \to \R_{\mbox{(\footnotesize mod 4)}} \to
 \R_{\mbox{(\footnotesize mod 4)}}\,{\sf X}\,\R}$
&$\;\;\Rightarrow\;\;$&
$\overbrace{Z_4 \,\to\, S^1 \,\to\, S^1\,{\sf  X R}}$\\
& &
\end{tabular}
\caption{Topology change of groups.}
\label{fig:D_group_topo}
\end{center}
\end{figure}

In general, the topology of such analytically extended groups progresses
from sets of points on a circle $S^1$ for discrete groups
generated by $d/dx$, to a circle $S^1$ for Lie groups
generated by $D^{\sigma'}_{\!x}$, and to a 2-dimensional cylinder
$\,S^1{\sf X R}\,$ for Lie group generated by $D^{\sigma'+it'}_{\!x}$ as
illustrated in Figure \ref{fig:D_group_topo}.

\section{Open Problems}

\begin{enumerate}
  
\item Develop the full theory of the symmetry breaking/deforming of
  the exact symmetry or gauge groups (e.g., Eightfold way $SU(3)$ and
  GUT $SU(5)$) to pseudo-groups with approximate symmetries that will
  fit the phenomenological data better.

\item As a mathematical curiosity, establish whether a meaningful
  extension between groups of the same family but of different
  dimensions, e.g., $SU(\Z) \mapsto SU(\R) \mapsto SU(\C)$ is
  possible.

\end{enumerate}

\end{document}